\documentclass[prl,aps,twocolumn,superscriptaddress,amsmath,amssymb]{revtex4}  
\usepackage{amsmath,epsfig}  
  
\begin{document}  
%
%
%
%
\title{Exact results for strongly-correlated fermions in 2+1 dimensions}
  
\author{Paul Fendley}
\affiliation{Department of Physics, University of Virginia,   
Charlottesville, VA 22904-4714 USA}

\author{Kareljan Schoutens}
\affiliation{Institute for Theoretical Physics, University of Amsterdam,
Valckenierstraat 65, 1018 XE Amsterdam, The Netherlands}
  
\date{May 24, 2005}

\begin{abstract}  

We derive exact results for a model of strongly-interacting spinless
fermions hopping on a two-dimensional lattice. By exploiting supersymmetry,
we find the number and type of ground states exactly.
Exploring various lattices and limits, we show how the ground states
can be frustrated, quantum critical, or combine frustration with a
Wigner crystal.  We show that on generic lattices, the model is in an
exotic ``super-frustrated'' state characterized by an extensive
ground-state entropy.

\end{abstract}  
  

\maketitle  

Over the last few decades thousands of papers have been written
exploring properties of itinerant-electron models in two spatial
dimensions. Exact results, however, for such systems at strong
coupling are few and far between. In this paper we find the exact
number and type of ground states in a model of spinless fermions with
strongly-repulsive nearest- and next-nearest-neighbor interactions. 
The strengths of these interactions are tuned to give an
exact {\em supersymmetry}. The supersymmetry not only makes our exact
computations possible, but balances competing terms in the
Hamiltonian. On most lattices, this results in an exotic
``super-frustrated'' state.

Our model is most transparently defined in terms of the supersymmetry
generator $Q$ and its hermitian conjugate $Q^\dagger$, which are
fermionic and obey $Q^2=(Q^\dagger)^2=0$. These commute with the
Hamiltonian defined by $H=\{Q,Q^\dagger\}$. This relation is at the heart of
supersymmetric quantum mechanics; a
number of important results follow \cite{Witten}.  All energy
eigenvalues $E$ satisfy $E\ge 0$, because $\langle s|H|s\rangle=
\langle s|QQ^\dagger|s\rangle+\langle s|Q^\dagger Q|s\rangle$ cannot
be negative.  Any state with $E=0$ is therefore a ground state; it is
annihilated by both $Q$ and $Q^\dagger$. 
Therefore all we need to do to construct a many-body model with
supersymmetry is to find a fermionic operator $Q$ squaring to zero.

Our degrees of freedom are spinless fermions living on any lattice or
graph of $N$ sites in any dimension. A fermion at site $i$ is created
by the operator $c_i^\dagger$ with $\{c_i,c^\dagger_j\}=\delta_{ij}$.
The sum $\sum_i c_i^\dagger$ squares to zero, but using this for $Q$
results in a trivial Hamiltonian. The strongly-interacting model we
will discuss was introduced in ref.\ \onlinecite{FSd}. The fermions
have a {\it hard core}, meaning that they are not only forbidden to be
on the same site as required by Fermi statistics, but are also
forbidden to be on adjacent sites. Their creation operator is
$d_i^\dagger=c_i {\cal P}_{<i>}$, where
\begin{equation}  
{\cal P}_{<i>}= \prod_{j\hbox{ next to } i} (1-c^\dagger_j c_j) 
\label{proj}
\end{equation}  
is zero if any site next to $i$ is occupied. 
A fermionic
operator $Q$ squaring to zero is then
$Q=\sum_i d^\dagger_i$.
This gives a non-trivial Hamiltonian
\begin{equation}
H=\{Q,Q^\dagger\}= \sum_{<ij>} d^\dagger_i d_j +  \sum_i {\cal P}_{<i>}.
\label{ham}  
\end{equation}
The latter term has a more conventional form 
on a lattice where every site has $z$
nearest neighbors: 
\begin{equation}
\sum_i {\cal P}_{<i>}=N-zF+\sum_{i}{V}_{<i>}
\label{potential}
\end{equation}
where ${V}_{<i>}+1$ is the number of particles adjacent to $i$,
unless there are none, in which case ${V}_{<i>}=0$. 
The operator $F=\sum_i
d^\dagger_i d_i$ counts the number of fermions. So in
addition to the hard core, the
Hamiltonian includes a hopping term, a constant (which we keep to
ensure ground states have $E=0$), a chemical potential $z$, and 
repulsive interactions between fermions two sites apart.

We use two mathematical
tools to study the $E=0$ ground states of (\ref{ham}). The first is
the {\em Witten index} $W$ \cite{Witten}. It is similar to the partition function, but includes
a minus sign for each fermion:
\begin{equation}
W = \hbox{tr}\left[(-1)^F e^{-\beta H}\right].
\label{Windex}
\end{equation}
$W$ is a lower bound on the number of ground states: it is the
difference of the number of bosonic ground states and the number of
fermionic ground states.  This is because all energy eigenstates with
$E>0$ form boson/fermion doublets of the same energy $E$ but opposite
$(-1)^F$. The states in a doublet contribute to $W$
with opposite signs and cancel, leaving only the sum of $(-1)^F$ 
over the ground states.

This argument shows that $W$ is independent of $\beta$, so we
can evaluate it in the $\beta \to 0$ limit, where every state
contributes with weight $(-1)^F$. We compute this by dividing the
lattice into two sublattices $S_1$ and $S_2$; we fix a configuration
on $S_1$, and sum $(-1)^F$ for the configurations on $S_2$. Then we sum the
results over the configurations on $S_1$. For a periodic
chain with $N=3j$ sites, we take $S_2$ to be every third site, and the
remaining sites $S_1$. Then the sum over configurations on any site on
$S_2$ vanishes unless at least one of the adjacent sites on $S_1$ is
occupied.
There are only two such configurations:
\begin{eqnarray}\nonumber
|\alpha\rangle&\equiv&\dots 
\bullet\square\circ\bullet\square\circ\bullet\square\circ\bullet\square\circ\bullet\square\circ\bullet\square\circ\bullet\square\circ\dots
\\ |\gamma\rangle&\equiv&
\dots\circ\square\bullet\circ\square\bullet\circ\square\bullet\circ\square\bullet\circ\square\bullet\circ\square\bullet\circ\square\bullet\dots
\label{alphagamma}
\end{eqnarray}
where the square represents an empty site on $S_2$. Both
$|\alpha\rangle$ and $|\gamma\rangle$ have $f=N/3$, so $W=2(-1)^f$, requiring
that are at least two ground states.

The second tool we use is the computation of the {\it cohomology} $H_Q$ of the
operator $Q$. This tool is even more powerful, allowing us to obtain
not just a lower bound, but rather the precise number of ground
states, and the fermion number of each.  The cohomology is the vector
space of states which are annihilated by $Q$ but which are not $Q$ of
something else (in mathematical parlance, these states are closed but
not exact) \cite{botttu}. Since $Q^2=0$, any state which is $Q$ of
something is annihilated by $Q$. Two states $|s_1\rangle$ and
$|s_2\rangle$ are said to be in the same cohomology class if
$|s_1\rangle = |s_2\rangle +Q|s_3\rangle$ for some state
$|s_3\rangle$.

The non-trivial
cohomology classes are in one-to-one correspondence with the $E$=0
ground states \cite{FSd}. To see this, consider an energy
eigenstate $|E\rangle$ with eigenvalue $E>0$. If $Q|E\rangle\ne 0$,
then it is not in any cohomology class. If $Q|E\rangle$=$0$
but $H|E\rangle \ne 0$, then $|E\rangle =
Q(Q^\dagger|E\rangle/E)$. This is in the trivial cohomology
class, so only the $E$=0 ground states have non-trivial
cohomology. Because they are annihilated by both $Q$ and
$Q^\dagger$, linearly independent $E$=0 ground states must be in different
cohomology classes. Precisely, the dimension of the vector space of
ground states (the ``number'' of ground states) is the same as that of
the cohomology. Since $F$ commutes with the
Hamiltonian, the cohomology class and the corresponding ground state
have the same fermion number.

To illustrate these techniques, let us first generalize some of the
one-dimensional results of ref.\ \onlinecite{FSd} to a staggered (but
still supersymmetric) chain. Let $Q(a)=Q_1 + aQ_2$ where $a$ is a
parameter and
\begin{equation}
Q_1=\sum_{j=1}^{N/3}
\left[d^\dagger_{3j-2} + d^\dagger_{3j}\right],\qquad 
Q_2=\sum_{j=1}^{N/3} d^\dagger_{3j-1}.
\label{q1q2}
\end{equation}
Because $[Q(a)]^2=0$,
the Hamiltonian $\{Q(a),Q^\dagger(a)\}$ is
supersymmetric. It deforms (\ref{ham}) by multiplying the hopping
term by $a$ for hopping on or off $S_2$, and multiplying ${\cal
P}_{\langle i\rangle}$ by $a^2$ when $i$ is on $S_2$. For
$a\to\infty$, the $E=0$ ground states therefore are the states
where $P_{<3j>}=0$ for all $j$. There are only two:
$|\alpha\rangle$ and $|\gamma\rangle$ from (\ref{alphagamma}). 
For large but finite $a$, there remain two ground states; for example
$|\alpha\rangle - 1/a|\alpha_2\rangle + {\cal O}(1/a^2)$, 
where $|\alpha_2\rangle$ is the sum of configurations
differing from $|\alpha\rangle$ by shifting one particle one site to
the right. When $a=1$, we know
from the Bethe ansatz solution that there are two ground states as
well, and that the model is gapless \cite{FNS}. For $a\ll 1$, there
are also two ground states, one localized on $S_2$, the other with one
particle for every two sites of $S_1$. The $a\ll 1$ ground states
spontaneously break different parity symmetries than $|\alpha\rangle$
and $|\gamma\rangle$ do, and $a=1$ is a quantum
critical point separating the two phases.

We find the exact number of ground states by computing the cohomology
$H_Q$ by using a {\em spectral sequence}. A useful theorem is the
``tic-tac-toe'' lemma of ref.\ \onlinecite{botttu}. This says that
under certain conditions, the cohomology $H_Q$ for $Q=Q_1+Q_2$ is the
same as the cohomology of $Q_1$ acting on the cohomology of $Q_2$.  In
an equation, $H_Q=H_{Q_1}(H_{Q_2})\equiv H_{12}$. As with our
computation of $W$, $H_{12}$ is found by first
fixing the configuration on all sites on the sublattice $S_1$, and
computing the cohomology $H_{Q_2}$. Then one computes the cohomology of
$Q_1$, acting not on the full space of states, but only on the classes
in $H_{Q_2}$. A sufficient condition for the
lemma to hold is that all non-trivial elements of $H_{12}$
have the same $f_2$ (the fermion number on $S_2$).

We apply this theorem to the one-dimensional chain by using the
decomposition of $Q=Q_1+aQ_2$ given by (\ref{q1q2}). Consider a single
site on $S_2$. If both of the adjacent $S_1$ sites are empty,
$H_{Q_2}$ is trivial: $Q_2$ acting on the empty site does not vanish,
while the filled site is $Q_2$ acting on the empty site. Thus
$H_{Q_2}$ is non-trivial only when every site on $S_2$ is forced to be
empty by being adjacent to an occupied site. The elements of $H_{Q_2}$
are just the two states $|\alpha\rangle$ and $|\gamma\rangle$ pictured
above in (\ref{alphagamma}).  Both states $|\alpha\rangle$ and
$|\gamma\rangle$ belong to $H_{12}$: they are closed because
$Q_1|\alpha\rangle =Q_1|\gamma\rangle=0$, and not exact because there
are no elements of $H_{Q_2}$ with $f_1=f-1$.  By the tic-tac-toe
lemma, there must be precisely two different cohomology classes in
$H_Q$, and therefore exactly two ground states with $f=N/3$. Applying
the same arguments to the periodic chain with $3f\pm 1$ sites and to
the open chain yields in all cases exactly one $E=0$ ground state,
except in open chains with $3f+1$ sites, where there are none
\cite{FNS}.

We emphasize that for finite $a$, $|\alpha\rangle$ and
$|\gamma\rangle$ are not the ground states themselves. A
representative of a cohomology class is not necessarily unique,
because adding $Q$ of something to it does not change the
class. A ground state is the one element in each class which
is also annihilated by $Q^\dagger$. One can use this observation in
principle (and in practice for small numbers of sites) to construct
the exact ground states from $|\alpha\rangle$ and $|\gamma\rangle$ as
a power series in $1/a$ \cite{angelis}.  The presence of states
$|\alpha\rangle$ and $|\gamma\rangle$ in the ground states hints that
the energy is lowest when particles are three sites apart. The
chemical potential favors the creation of more particles, but putting
them two sites apart causes an increase in potential energy and
hopping energy. The two effects balance at an average separation of
roughly three sites; we call this heuristic the ``3-rule''.

Having introduced the mathematical tools necessary, we now turn to the
study of our spinless-fermion model on two-dimensional
lattices. We find that generically, there is an {\em extensive ground
state entropy}: the number of ground states increases exponentially
with the size of the system. This indicates that the system is
frustrated; we will explain how in the following.

The systematics of the one-dimensional case quickly extend to lattices
of type $\Lambda_3$, which are obtained from any lattice (or even
graph) $\Lambda$ by putting two additional sites on every
link. Letting $S_1$ be the original sites of $\Lambda$ and $S_2$ the
added sites, the only states in $H_{Q_2}$ and $H_{12}$ are the two
where $S_1$ is completely full, and completely empty. The first gives
an $E=0$ ground state with $f=N_\Lambda$ (the number of sites of
$\Lambda$), while the latter gives an $E=0$ state with $f=L_\Lambda$
(the number of links in $\Lambda$), with a possible exception when
$L_\Lambda=N_\Lambda- 1$.  When $\Lambda$ is the square lattice, the
two ground states on $\Lambda_3$ have filling $f=N/5$ and $f=2N/5$.
Lattices of type $\Lambda_3$ are the only two-dimensional ones we know
of where the number of ground states does not grow with the size of
the lattice.

\begin{figure}[h!] 
\begin{center} 
\includegraphics[width= .45\textwidth]{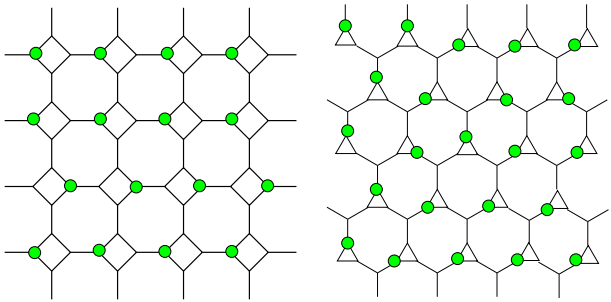} 
\caption{Configurations obeying the 3-rule on the octagon-square and
nonagon-triangle lattices}
\label{fig:octagon} 
\end{center} 
\end{figure}

Another exceptional case is the octagon-square lattice
in the first part of fig.\ \ref{fig:octagon}.  We take $L$
rows and $M$ columns of squares (hence $N=4LM$ sites).  Let $S_1$
consist of the leftmost site on every square.  Then $H_{Q_2}$ is
trivial unless all the $M$ sites on $S_1$ in a given row either all
are occupied, or all are empty. There are $2^L-1$ such configurations
which have at least one row in $S_1$ occupied. Because of the hard
core, all the sites of $S_2$ adjacent to an occupied site on $S_1$
cannot be filled, and the remaining sites form independent open chains
of length a multiple of 3. Such an open chain has just one element of
$H_{Q_2}$, so each of these $2^L-1$ configurations correspond to one
element of $H_{Q_2}$ and $H_{12}$. Now consider the configuration
where all sites on $S_1$ are empty, so that the sites on $S_2$ form
$M$ periodic chains, each of length $3L$.  We showed above that
$H_{Q_2}$ for {\em each} of these chains has {\em two} independent
elements. Thus $H_{Q_2}$ and $H_{12}$ are of dimension $2^L+2^M-1$.
Applying the tic-tac-toe lemma to this case is more involved, 
but the conclusion is that there are $2^L+2^M-1$ ground
states, each with $N/4$ fermions.

We believe that on the octagon-square lattice, the model exhibits a
combination of Wigner-crystal order with frustration. There are
$2^L+2^M$ configurations of $N/4$ particles which satisfy our
heuristic 3-rule. $2^L$ of them are of the form displayed in fig.\
\ref{fig:octagon}: one can shift all the particles in a given row
without violating the rule. This illustrates how frustration
arises: in each row one can shift all the particles without violating
the 3-rule. Likewise, $2^M$ of them have particles on
the top or bottom of each square. For mysterious reasons, the state
with $(k_x,k_y)$=0 is not a ground state, but we believe the remaining
$2^L+2^M-1$ ordered states dominate the actual ground states.
In further support of this claim, we analyze the 
discrete symmetries commuting with $Q$. If a
given element of the cohomology spontaneously breaks such a symmetry,
the corresponding ground state will break it too. The ground states
have spontaneously-broken parity symmetries like the Wigner crystal
states in fig.\ \ref{fig:octagon}. Again like the crystal, all but one
of the $2^L-1$ ground states first considered spontaneously break
translation symmetry in the vertical direction but not the horizontal;
$2^M-2$ of the remaining ground states spontaneously break translation
symmetry in the horizontal direction.  Moreover, the number of ground
states here can be changed by requiring that just one site anywhere on
the lattice be occupied. Consider the octagon-square lattice with one
site on $S_1$ and its three neighbors on $S_2$ removed; this is
equivalent to demanding that there be a particle on this $S_1$
site. On this lattice there are just $2^{L-1}$ ground states. Only in
an ordered system should this type of change occur.

The $\Lambda_3$ and octagon-square lattices are exceptional: on all
other lattices we have studied the ground-state
en\-tro\-py is extensive.  In many cases (including the triangular, hexagonal and
Kagom\'e lattices), this can be seen by computing the Witten
index $W$ as a function of the size of the lattice.  Employing a
row-to-row transfer matrix $T_M$, the index for $M\times L$ unit cells
is expressed as $W_{L,M}= {\rm tr} [ (T_M)^L ]$. We found by exact
diagonalization that the largest eigenvalues
$\lambda_M^{\rm max}$ of the $T_M$ here behave as $\lambda_M^{\rm
max}\propto \lambda^M$, with $|\lambda|>1$.  Clearly, the absolute
value $|\lambda|$ sets a lower bound on the ground-state entropy per
lattice site. For $n$ sites per unit cell,
$ S_{\rm GS}/N \geq \ln |W_{L,M}| /(nML) \sim \ln |\lambda|/n$.
For the triangular lattice, 
$S_{\rm GS}/N \geq 0.13$. \cite{Hendrik,FSv} 

For the nonagon-triangle lattice shown in the right half of fig.\
\ref{fig:octagon}, the extensive ground-state entropy can be exactly
computed.  This lattice is formed by replacing every
other site on a hexagonal lattice with a triangle. To find the ground
states, take $S_1$ to be the sites on the triangles, and $S_2$ to be
the remaining sites. As with the chain, $H_{Q_2}$ vanishes unless
every site in $S_2$ is adjacent to an occupied site on some triangle.
The non-trivial elements of $H_{Q_2}$ therefore must have
precisely one particle per triangle, each adjacent to a different site
on $S_2$. This is because a triangle can have at most one particle on
it, and (with appropriate boundary conditions) there are the same
number of triangles as there are sites on $S_2$.  A typical element of
$H_{Q_2}$ is shown in fig.\ \ref{fig:octagon}. One can think of these
as ``dimer'' configurations on the original honeycomb lattice, where
the dimer stretches from the site replaced by the triangle to the
adjacent non-triangle site. Each close-packed hard-core dimer
configuration is in $H_{12}$, and by the tic-tac-toe lemma, it
corresponds to a ground state. The number of such ground states
$e^{S_{\rm GS}}$ is therefore equal to the number of such dimer coverings
of the honeycomb lattice, which for large $N$ is \cite{dimer}
\begin{equation}
\frac{S_{\rm GS}}{N} =\frac{1}{\pi}
\int_0^{\pi/3}d\theta\ln[2\cos(\theta)] = 0.16153\ldots
\end{equation}
The frustration here clearly arises because there are many
ways of satisfying the 3-rule.

For the staggered model, $Q=Q_1+aQ_2$, on the non\-a\-gon-triangle lattice,
the dimer states are the exact ground states when $a\to\infty$.  In
the singular limit $a=0$ there are more ground states:
$|\Psi_2\rangle$ with all particles on $S_2$; $2^{N/4}$ ground states
$|\Psi_1^{(s)}\rangle$ with one particle on each of the triangles; and
additional ground states at higher fermion numbers as well.  For
$0<a\ll 1$, $|\Psi_2\rangle$ and $e^{S_{\rm GS}}-1$ of the
$|\Psi_1^{(s)}\rangle$ remain ground states,
while the others develop energies of order $a^2$.
The ground-state degeneracy can be lifted by including terms that
break the supersymmetry. Consider changing the
intra-triangle hopping amplitude to $1-\epsilon$ with $\epsilon>0$.
At $a=0$, the
 $|\Psi_1^{(s)}\rangle$ have energy $E=N\epsilon/4$, so here
the Wigner
crystal $|\Psi_2\rangle$ is the unique $E$=0 ground state.  For $a$ large and
$\epsilon$ small, the leading piece in the effective Hamiltonian is a
``flip'' of 3 dimers around a plaquette, as in the quantum dimer model
\cite{RK}. For generic potentials this model orders \cite{MSC},
leading to the possibility of a quantum critical point intermediate
between this ordered phase and the $a=0$ one. A quantum critical
point indeed occurs for
the chain at $a=1$ [\onlinecite{FSd},\onlinecite{FNS}], and so seems possible
on general lattices for $a \sim 1$.

The situation on lattices with higher coordination number is more complicated. 
There are ground states with more particles than the 3-rule allows:
the increased chemical potential and possibilities for hopping
compensate for an increase in potential energy. 
\begin{figure}[ht] 
\begin{center} 
\includegraphics[width= .40\textwidth]{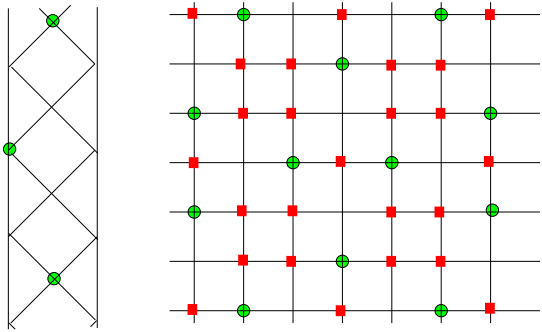} 
\caption{``Ordered'' states for the triangle-square ladder and the square
lattice; the red squares are sublattice 2}
\label{fig:square2} 
\end{center} 
\end{figure} 
For the triangle-square ladder 
in fig.\ \ref{fig:square2}
with $N=3n+1$ sites and open boundary conditions, 
we obtained a recursion relation for
the ground-state generating function 
$P_n(z)={\rm tr}_{\rm GS}(z^F)$:
\begin{equation}
P_{n+3}(z)=2 z^2 P_n(z) + z^3 P_{n-1}(z) \ ,
\end{equation}
with $P_0=0$, $P_1=z$, $P_2=2z^2$, $P_3=z^3$. This shows the existence
of $2^{n/3}$ ground states at fermion number $2N/9$, and also
indicates additional ground states at higher fillings, up to $N/4$. An
``ordered'' state with $f=2N/9$ violating the
3-rule is given in fig.\ \ref{fig:square2}, but the
frustration is evident in that there are many such states. Using the recursion relation, we find
that the ground-state entropy is set by
the largest solution $\lambda^{\rm max}$ of $\lambda^4-2\lambda-1=0$,
giving $S_{\rm GS}/N = (\ln \lambda^{\rm max})/3 = 0.1110\dots$.

On a square lattice of $3L\times 3M$ sites with periodic boundary
conditions, the situation is similar. When $S_2$ consists of the red
squares in fig.\ \ref{fig:square2}, there are two elements of $H_{12}$
which have $S_2$ empty; one of them is displayed in fig.\
\ref{fig:square2}.  They have $2N/9=2LM$ particles, and also violate
the 3-rule.  Many more ground states with different
fermion numbers can be found by introducing various types of defects
in this pattern, but we have not found a way of counting them all
\cite{FSv}.

Our exact results indicate that there is a new kind of exotic phase
for itinerant fermions on a two-dimensional lattice with strong
interactions.  This ``super-frustrated'' state exhibits an extensive
ground-state entropy, and occurs because supersymmetry ensures a
perfect balance between competing terms in the Hamiltonian. Patterns
with charge order can be distinguished in various limits and on
special lattices, but the effect of (approximate) supersymmetry in
general is that defects between different domains come at zero (very
low) energy cost. For example, the charge order (stripes) found for
hard-core fermions on the square lattice \cite{ZH} becomes
super-frustrated as the interactions and chemical potential are tuned
to the supersymmetric point.

{\bf Acknowledgments:} We thank J.~de Boer and B. Nien\-huis for
collaboration and discussions, H.~van Eerten for sharing his
unpublished numerical results, and P. Arnold for coining
``super-frustrated''. This work was supported by the foundations FOM
and NWO of the Netherlands, by the NSF via the grant DMR-0412956, and
by the DOE under grant DEFG02-97ER41027.

\end{document}